\documentclass[nofootinbib,aps,preprintnumbers,unsortedaddress,prl,twocolumn]{revtex4}
\usepackage{graphicx}
\usepackage{amssymb}
\usepackage{textcomp}
\usepackage{amsmath}
\usepackage{bm}
\usepackage{times}
\usepackage{graphics}
\usepackage{hyperref}
\usepackage{setspace}
\usepackage{paralist}
\usepackage{amsmath}

\usepackage{soul}

\usepackage{slashed}
\usepackage{comment}
\usepackage{array}

\usepackage{rotating}
\usepackage{framed}

\usepackage{placeins}

\usepackage{amsmath} 
\usepackage{amsthm} 
\usepackage{amssymb}	
 
\newcommand{\avg}[1]{\left< #1 \right>} 
 
\let\baraccent=\= 
\renewcommand{\=}[1]{\stackrel{#1}{=}} 

\theoremstyle{definition}

\theoremstyle{remark}

\usepackage{bbm}

\hypersetup{
    pdfnewwindow=true,      
    colorlinks=true,       
    linkcolor=black,          
    citecolor=blue,        
    filecolor=blue,      
    urlcolor=blue           
}

\makeatletter
\newcommand\xleftrightarrow[2][]{%
  \ext@arrow 9999{\longleftrightarrowfill@}{#1}{#2}}
\newcommand\longleftrightarrowfill@{%
  \arrowfill@\leftarrow\relbar\rightarrow}
\makeatother

\begin{document}
\title{\Large {\it{\bf{Sterile Neutrinos and B-L Symmetry}}}}
\author{Pavel Fileviez P\'erez$^{1}$, Clara Murgui$^{2}$}
\affiliation{$^{1}$Department of Physics, Center for Education and Research in Cosmology and Astrophysics (CERCA), 
and Institute for the Science of Origins (ISO), Case Western Reserve University,
Rockefeller Bldg. 2076 Adelbert Rd. Cleveland, OH 44106, USA \\
$^{2}$Departamento de F\'isica Te\'orica, IFIC, Universitat de Valencia-CSIC, 
E-46071, Valencia, Spain}
%
%
\begin{abstract}
We revisit the relation between the neutrino masses and the spontaneous breaking of the $B-L$ gauge symmetry.
We discuss the main scenarios for Dirac and Majorana neutrinos and point out two simple mechanisms for neutrino masses. 
In this context the neutrino masses can be generated either at tree level or at quantum level and one predicts the existence of very light sterile neutrinos with masses below the eV scale.
The predictions for lepton number violating processes such as $\mu \to e$ and $\mu \to e \gamma$ are discussed in detail. The impact from the cosmological constraints on the effective number of relativistic degree of freedom is investigated.
\end{abstract}
\maketitle 
\section{I. Introduction}
The discovery of the Standard Model (SM) boson responsible for the electroweak 
symmetry breaking five years ago was crucial to establish the SM as one of the 
successful theories of nature. Nowadays it is well-known the mechanism responsible to 
generate masses for the charged fermions in the SM but 
unfortunately we cannot explain the ratio between their masses.

Today, we know that the neutrinos are not massless: the solar and atmospheric 
mass squared differences are known from neutrino experiments with a very good precision, 
see Ref.~\cite{Esteban:2016qun} for the current values, and there are some important 
bounds from cosmology, see for example Refs.~\cite{Ade:2015xua,Palanque-Delabrouille:2015pga,Cuesta:2015iho,Giusarma:2016phn,Hannestad:2016fog}. However, we still do not have any clue about the mechanism behind their mass generation. Clearly, the fact that in the Standard Model neutrinos are exactly massless forces one to go beyond to understand the origin of their masses.
See Refs.~\cite{Mohapatra:2006gs,Cai:2017jrq,FileviezPerez:2009ym} for recent reviews about neutrino mass mechanisms.

In the Standard Model the charged fermion masses are proportional to the electroweak (EW) symmetry breaking scale.
In the case of the neutrinos, the simplest way one can relate their masses to a new symmetry breaking scale is 
to consider scenarios where $B-L$ is a local gauge symmetry. Here $B$ and $L$ are for baryon and lepton numbers, respectively.
As it is well-known, $B-L$ can be an anomaly free local symmetry by adding three copies of right-handed neutrinos to the standard fermion content.
If $B-L$ is never broken one can explain why neutrinos are Dirac particles, while when it is spontaneously broken one can 
investigate the generation of Majorana masses.

In this article we revisit the connection between the neutrino masses and the $B-L$ symmetry breaking scale.
We discuss the different scenarios where the neutrinos can be Dirac or Majorana fermions. 
In the case where they are Dirac fermions, we discuss the $B-L$ Stueckelberg extension of the SM.
We also discuss the well-known scenario of canonical seesaw, where the $B-L$ symmetry is spontaneously broken in two 
units. In this context, the right-handed neutrinos are typically heavy and the light neutrinos are Majorana particles.
However, in this letter, we point out two scenarios where the neutrinos are Majorana particles and one predicts the existence of very 
light right-handed neutrinos. In the first scenario, the $B-L$ is broken in two units but the right-handed neutrinos are very light, with masses 
below the eV scale. In this case, the neutrino masses are generated at tree level through the inverse Type II `seesaw' mechanism.
In the second mechanism, the right-handed and left-handed Majorana neutrino masses are generated at the one-loop level. 
In this case the right-handed neutrinos are also very light. 

We investigate the main phenomenological constraints for the neutrino mass mechanisms where the right-handed neutrinos are very light.
We discuss in detail the cosmological bounds on the effective number of relativistic degrees of freedom to impose non-trivial bounds 
on the neutrino interactions. We show that the cosmological bounds are as competitive as the current collider bounds 
on new gauge bosons interacting with all the SM fermions. The predictions for the lepton number violating processes such 
as $\mu \to e$ and $\mu \to e \gamma$ are investigated in detail. Lepton flavor violating (LFV) transition searches are nowadays one of the most sensitive probes of new physics and their sensitivity is expected to be improved at least 3-4 orders of magnitude in the near future. Therefore, the prediction of light sterile neutrinos and testable lepton number violating signals at the current and future experiments make the Radiative seesaw model proposed in this Letter an appealing mechanism to generate neutrino masses in the context of $B-L$ gauge symmetries. 

This letter is organized as follows:  In section II we discuss the main mechanisms 
for neutrino masses in simple theories where $B-L$ is a local symmetry, 
in section III the main features of the $B-L$ radiative seesaw mechanism are discussed, 
in section IV we discuss the cosmological bounds, in section V we discuss the predictions 
for lepton number violating processes, while in section VI we summarize the main results.
\section{II. Neutrino Masses and B-L Gauge Symmetry}
It is very well-known that there is a simple connection between the generation of neutrino masses and the $B-L$ gauge symmetry.
The $B-L$ local symmetry is the simplest symmetry which can be anomaly free by adding three copies of right-handed neutrinos, i.e. $\nu_R^i$ with $i=1,2,3$.
Here we discuss the simplest mechanisms for neutrino masses where the $B-L$ gauge symmetry is spontaneously broken and define the seesaw scale.
\begin{itemize}

\item {\textit{Dirac Neutrinos}}: 
As the other SM fermions, neutrinos can be Dirac fermions, and in this case the relevant Lagrangian is given by 
\begin{equation} 
- {\cal L} \supset Y_\nu \ \overline{\ell}_L i \sigma_2 H^* \nu_R  \ + \  \textrm{h.c.},
\end{equation}
where $\ell_L \sim (1,2,-1/2,-1)$, $H\sim (1,2,1/2,0)$, and $\nu_R \sim (1,1,0,-1)$. Here the local $B-L$ gauge symmetry forbids the Majorana mass for right-handed neutrinos, 
and the gauge boson $Z_{BL}$ can acquire mass in two different ways:

a) Using the Stueckuelberg mechanism~\cite{Stueckelberg:1900zz} one can generate a mass for the  $Z_{BL}$ without breaking the gauge symmetry~\cite{Feldman:2011ms} through the following terms  
\begin{equation} 
- {\cal L}_{St} \supset \frac{1}{2} \left( M_{BL} Z^\mu_{BL} + \partial^\mu \sigma \right) \left( M_{BL} Z_{BL \mu} + \partial_\mu \sigma \right),
\end{equation}
where the gauge transformation is written as $\delta Z^\mu_{BL} = \partial^\mu \lambda$ and $\delta \sigma = - M_{BL} \lambda$.
See Ref.~\cite{Feldman:2011ms} for a detailed study.

b) One can break $B-L$ through the Higgs mechanism where 
the new Higgs $S_{BL}$ has a $B-L$ quantum number larger than two, and with the minimal field content one cannot generate Majorana masses for the right-handed neutrinos. 

In both cases the neutrino masses are given by $M_\nu = \frac{1}{\sqrt{2}} Y_\nu v_H$, with $v_H = \sqrt{2} \left<H^0\right>=246$ GeV and $Y_\nu \approx 10^{-13}-10^{-12}$ in order to reproduce the values 
of the squared mass differences measured in the neutrino experiments. 

\item {\textit{Canonical Seesaw}}~\cite{TypeI}: 
In the case when $S_{BL} \sim (1,1,0,2)$ breaks $B-L$ one can generate Majorana masses for the right-handed neutrinos at tree level. 
This is the case of canonical Type I seesaw and the relevant Lagrangian is given by
\begin{equation}
-{\cal L}_\nu^I = Y_\nu \ \overline{\ell_L} i \sigma_2 H^* \nu_R + \lambda_R \ \nu_R^T C \nu_R S_{BL} + \text{h.c.},
\end{equation}
with $\lambda_R = \lambda^T_R$. The neutrino mass matrix in the basis $(\nu, \  \nu^C)$ reads as 
\begin{equation}
 {\cal M}_\nu^I = \begin{pmatrix} 0 && M_\nu^D \\ (M_\nu^D)^T && M_\nu^R \end{pmatrix},
\end{equation}
where 
\begin{eqnarray}
M_\nu^D&=& \frac{1}{\sqrt{2}}Y_\nu v_H,\ {\text{and}} \
M_\nu^R= \sqrt{2} \ \lambda_R v_{BL}.
\end{eqnarray}
Here $v_{BL}=\sqrt{2} \left< S_{BL} \right>$ defines the seesaw scale.
In this case, the right-handed neutrino masses can be large and the upper bound on the $B-L$ breaking scale is around $10^{14}$ GeV.
Therefore, there is a priori no reason to expect this particular realization of the seesaw mechanism to be tested in the near future. 

In the case when the right-handed neutrino masses are below the TeV scale, they can be produced through the $B-L$ gauge boson, i.e. $pp \to Z_{BL}^* \to NN$, see 
for example Ref.~\cite{Perez:2009mu} for the study of these signatures at the LHC. 
It is important to emphasize that in the context of the canonical seesaw the symmetry breaking scale can be large 
and we might never be able to test this idea. 

\item {\textit{$B-L$ Inverse Type II seesaw}}: 
One can have a different scenario for the generation of neutrino masses by breaking the $B-L$ symmetry with a scalar triplet $\Delta \sim (1,3,1,2)$, which generates Majorana masses for the left-handed neutrinos. In this context, the $B-L$ symmetry is broken in two units but the right-handed neutrinos are very light as we will show.
The relevant Lagrangian for our discussion is given by
\begin{equation}
- {\cal L}_\nu^{II} = Y_\nu \overline{\ell_L} i\sigma_2 H^* \nu_R + \lambda_L \ell_L^T C i\sigma_2 \Delta \ell_L + \text{h.c.}, 
\end{equation}
with $\lambda_L=\lambda_L^T$ and $\Delta$ is given by
\begin{equation}
\Delta = \begin{pmatrix} \delta^+/ \sqrt{2} && \delta^{++} \\ \delta^0 && - \delta^+/\sqrt{2} \end{pmatrix}.
\end{equation}
In this context the neutrino mass matrix in the basis $(\nu, \  \nu^C)$  reads as
\begin{equation}
{\cal M}_\nu^{II} = \begin{pmatrix} M_\nu^L && M_\nu^D \\ (M_\nu^D)^T && 0 \end{pmatrix},
\end{equation}
where
\begin{eqnarray}
M_\nu^L&=& \sqrt{2} \lambda_L v_\Delta,
\end{eqnarray} 
with $v_\Delta/\sqrt{2}$ being the vacuum expectation value of the neutral component of the triplet, $\delta^0$. 
Clearly, in this scenario the right-handed neutrino masses will be smaller or have similar values as the left-handed neutrino masses.  
In this case there are two main possibilities to consider: 
\begin{itemize}

\item Pseudo-Dirac neutrinos when $M_\nu^L << M_\nu^D$, 

\item Majorana neutrinos when $M_\nu^D << M_\nu^L$. 

\end{itemize}
In order to avoid large mixing between the active and sterile neutrinos one should work in the limit 
$M_\nu^D << M_\nu^L$, and in this case the neutrino masses are given by 
$$M_{\nu_L} \approx M_\nu^L, \qquad \textrm{and} \qquad M_{\nu_R} \approx (M_\nu^D)^2/M_\nu^L.$$  
Then, we have the interesting result that the right-handed neutrinos, `sterile' neutrinos, must be very light even if $B-L$ has been broken in two units. 

Now, since the vacuum expectation value of the $\Delta$ field cannot be large, $v_\Delta \lesssim 4 \text{ GeV}$, one needs to add a new Higgs, $S \sim (1,1,0,n_{BL})$ with $|n_{BL}| > 2$, in order to generate a large mass for the $B-L$ gauge boson. Here $|n_{BL}| > 2$ is required to avoid any higher-dimensional operator which could generate masses for the right-handed neutrinos. Unfortunately, in this case one predicts the existence of an extra Goldstone boson, the Majoron, and one has a new contribution to the $Z$ decays, $Z \to J \delta_R$. 
Here $J$ is for the massless Majoron and $\delta_R$ for the light CP-even Higgs. This model is ruled out as the original Roncadelli-Gelmini model~\cite{RG}.

It is important to mention that the simplest scenario, with only the scalar triplet and the SM Higgs in the scalar sector, can be realistic because the Majoron is eaten by the $Z_{BL}$. 
However, since $v_\Delta \lesssim 4 \text{ GeV}$ the $Z_{BL}$ has to be very light and one needs to assume a very small $g_{BL}$ gauge coupling to satisfy all experimental bounds, see for example~\cite{Batell:2016zod,Heeck:2014zfa}, and therefore it is very difficult or impossible to test this mechanism.
%
\item $B-L$ {\textit{Radiative Seesaw Mechanism:}}
\begin{figure}
\includegraphics[width=0.75\linewidth]{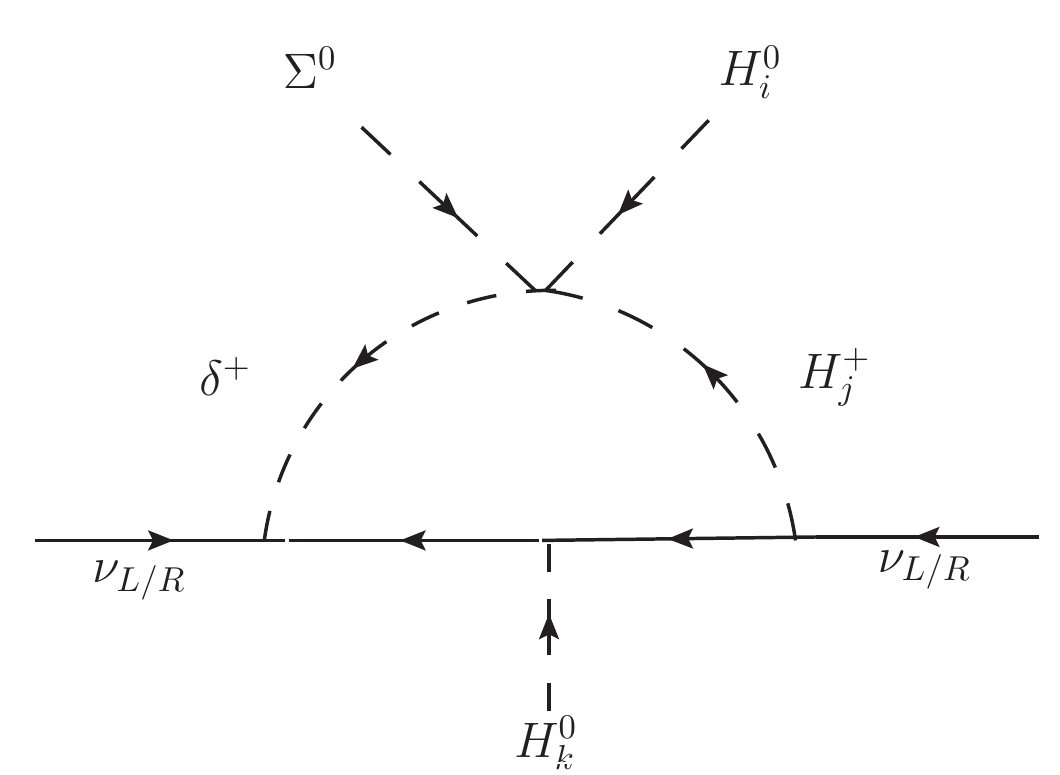}
\caption{$B-L$ Radiative Seesaw Mechanism in the unbroken phase.}
\end{figure}
Now, we would like to point out a second mechanism for neutrino masses where the $B-L$ symmetry is spontaneously broken. 
One can generate neutrino masses at one-loop level using the Zee-mechanism~\cite{Zee}.  In this scenario we study a simple extension of the Zee mechanism where the local $B-L$ gauge symmetry is spontaneously broken. In order to 
generate neutrino masses only through the Zee mechanism the needed interactions are given by
\begin{eqnarray}
- {\cal L}_\nu^{\text{RS}} &=& \lambda_L \ell_L^T C i \sigma_2 {\ell_L}\delta^+ + \lambda_R \nu_R^T C e_R \delta^+ \nonumber \\
&+& \lambda_{ij} H_i^T i \sigma_2 \Sigma H_j \delta^- + Y_e^i \bar{\ell}_L H_i e_R \nonumber \\
&+ & Y_\nu^i \bar{\ell}_L i \sigma_2 H_i^* \nu_R + \text{h.c.},
\end{eqnarray}
with $\lambda_L =- \lambda_L^T$, $i=1,2$, and the fields $\delta^+ \sim (1,1,1,2)$, $H_i \sim (1,2,1/2,0)$ and $\Sigma \sim (1,3,0,2)$ which is given by
\begin{equation}
\Sigma = \frac{1}{\sqrt{2}} \begin{pmatrix} \Sigma^0 && \sqrt{2} \Sigma^{+}_1 \\ \sqrt{2} \Sigma^-_2 && - \Sigma^0 \end{pmatrix}.
\end{equation}
In this case one can generate masses for the left and right-handed neutrinos at the one-loop level 
according to Fig. 1 and, as we will explain in the next sections, the right-handed neutrinos have to be light in this context.
In this case the neutrino masses, as in the previous scenario, are proportional to the vacuum expectation value of the real 
triplet breaking the local $B-L$ which has to be below the GeV scale. Notice that the field $\Sigma$ cannot generate 
masses for the right-handed neutrinos at tree level. As in the previous case, in order to generate a large mass for the 
$B-L$ gauge boson, a new Higgs, $S \sim (1,1,0,-4)$ must be included in this model.
\end{itemize}
As one can appreciate, we have pointed out two models based on the spontaneous breaking of the $B-L$ gauge symmetry 
where the right-handed neutrinos are very light with mass below the eV scale. Unfortunately, in the case of Inverse Type II seesaw 
one needs to assume a very small $B-L$ gauge coupling to be in agreement with the experiment. In the next section we will focus 
on the $B-L$ Radiative Seesaw Mechanism which can be realistic and could be tested in current or future experiments.
%
\section{III. $B-L$ Radiative Seesaw Mechanism}
%
As we have discussed before, the neutrino masses can be generated at one-loop level as we have shown in Fig.~1. In this scenario in order to generate neutrino masses one 
has two Higgs doublets (including the SM Higgs) $H_i \sim (1,2,1/2,0)$,  a singly charged Higgs $\delta^+\sim (1,1,1,2)$ and a Higgs triplet $\Sigma  \sim (1,3,0,2)$. 
Here we discuss some of the main features of this model. The W-mass in this case is given by
\begin{equation}
M_W^2=\frac{1}{4}g_2^2(v_1^2+v_2^2 + 4 v_\Sigma^2),
\end{equation}
with $v^2=v_1^2+v_2^2+4 v_\Sigma^2$. Here $v_i/\sqrt{2}$ is the vacuum expectation value of the Higgs doublet $H_i$. 

In this scenario there is no mixing between the new neutral gauge boson $Z_{BL}$ and the rest of SM gauge bosons. 
Since the vacuum expectation value of the triplet contributes to the W-mass, one finds that the variation of the $\rho$ parameter is given by 
\begin{equation}
\delta \rho = \rho -1 = \frac{M_W^2}{M_Z^2 \cos^2 \theta_W}-1= \frac{4 v_\Sigma^2}{v_1^2+v_2^2}.
\end{equation}
As in the case of the Inverse Type II seesaw, the $\rho$-parameter imposes an upper bound on the triplet vacuum expectation value, $v_\Sigma \lesssim 3$ GeV~\cite{FileviezPerez:2008bj}. 
In this context the mass of the new gauge boson is given by 
\begin{equation}
M_{Z_{BL}}^2=g_{BL}^2 (16 v_S^2 + 4 v_\Sigma^2).
\end{equation}
Here $v_S/\sqrt{2}$ is the vacuum expectation value of the field $S \sim (1,1,0,-4)$ needed to generate a large mass for the $B-L$ gauge boson. Here $S$ plays a twofold role:
since in the scalar potential the term $\textrm{Tr} \Sigma^2 S$ is allowed one avoids the existence of extra Goldstone bosons and since the vacuum expectation value 
can be large one can satisfy the experimental bounds on the $B-L$ gauge boson without assuming a small gauge coupling. 

Using the interactions in Eq.~(10) one can compute the neutrino masses generated at the one-loop level. The mass matrix for the charged Higgses is diagonalized by the following unitary matrix V,
\begin{equation}
\begin{pmatrix} H_1^+ \\ H_2^+ \\ \Sigma^+_1 \\ \Sigma^+_2 \\ \delta^+ \end{pmatrix} = V \begin{pmatrix}  h_1^+ \\ h_2^+ \\ h_3^+ \\ h_4^+  \\ h_5^+\end{pmatrix},
\end{equation}
and the mass matrix for neutrinos is given by
\begin{equation}
 {\cal M}_\nu= \begin{pmatrix}  M_\nu^L && M_\nu^D \\ (M_\nu^D)^T && M_\nu^R \end{pmatrix},
\end{equation}
where
\begin{eqnarray}
(M_\nu^L)^{\alpha \gamma}&=& \frac{1}{8\pi^2} \sum_\beta \lambda_L^{\alpha \beta}m_{e_\beta}\sum_i \text{Log}\left(\frac{m_{h_i}^2}{m_{e_\beta}^2}\right) \nonumber \\
&\times& ({Y_{e1}^\dagger}^{\beta \gamma}V_{1i}^*+{Y_{e2}^\dagger}^{\beta \gamma}V_{2i}^*)V_{5i} + \alpha \leftrightarrow \gamma,\\
(M_\nu^R)^{\alpha \gamma}&=&\frac{1}{(4\pi)^2} \sum_\beta \lambda_R^{\alpha \beta} m_{e_\beta}\sum_i \text{Log}\left(\frac{m_{h_i}^2}{m_{e_\beta}^2}\right) \nonumber \\
&\times& (Y_{\nu 1}^{\beta \gamma} V_{1i}^*+Y_{\nu 2}^{\beta \gamma}V_{2i}^*)V_{5i}+ \alpha \leftrightarrow \gamma.
\end{eqnarray}
In this case when $Y_\nu$ is very small one has an inverse seesaw for the neutrino masses since $M_\nu^L >> M_\nu^D, M_\nu^R$.
This scenario represents the most interesting case since one can have a small mixing angle between the left-handed and right-handed neutrinos.
Therefore, as in the case of the Inverse Type II seesaw mechanism, here we predict the existence of light right-handed neutrinos. Their masses 
should be below or at the scale of the left-handed neutrinos. 

In order to complete our discussions we show in Fig.~2 the branching ratios for the $B-L$ gauge boson for different mass values. 
As we can see in Fig.~2 the invisible branching ratio can be very large, between $40\%-30\%$ in the mass range shown, due to the 
presence of very light right-handed neutrinos. The branching ratio into charged leptons is basically equal to the invisible decays as we show in Fig.~2.
In this model, neglecting the mixing among the scalars without loss of generality, the $Z_{BL}$ can decay into singly charged Higgses, $\delta^\pm$ and $\Sigma_{1/2}^\pm$ in the triplet, as we have shown in Fig.~2. We do not consider here the possibility of $Z_{BL}$ decaying into neutral Higgses since the massive CP-odd field is predicted to be at the $B-L$ scale. 
Only for illustration we use the values $m_{\Sigma_1^{+}}=m_{\Sigma_2^{+}}=400$ GeV and $m_{\delta^{+}}=600$ GeV. In summary, the $B-L$ gauge boson has a large invisible 
branching ratio and the singly charged Higgses can be produced through this new force.
\begin{figure}[t]
\includegraphics[width=0.99\linewidth]{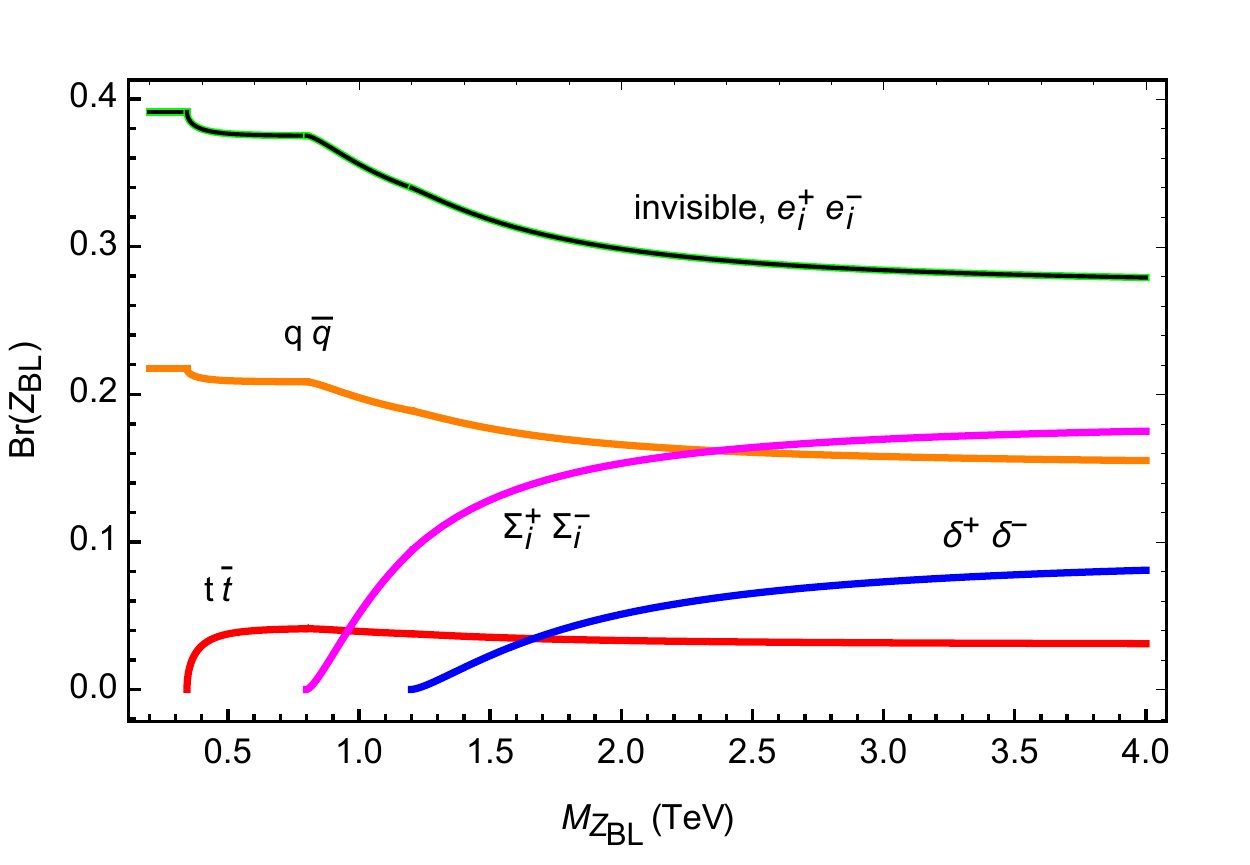}
\caption{Branching ratios for the $B-L$ Gauge Boson. The green line represents the invisible decays, which overlaps with the black line, corresponding to the decays into two charged leptons. The orange line corresponds to the decays into two quarks except for the decay into two top quarks, which is represented by the red line. The pink and blue lines correspond to the decays into two charged Higgses, $\Sigma_{1/2}^\pm$ from the triplet and $\delta^\pm$, respectively. Here we have neglected any mixing among scalars for simplicity. The values  $m_{\Sigma_1^+}=m_{\Sigma_2^+}=400$ GeV and $m_{\delta^+}=600$ GeV have been taken for illustration.}
\label{BrZBL}
\end{figure}
%
{\section{IV. Cosmological Bounds}}
%
In the two models for neutrino masses presented above one predicts the existence of very light right-handed neutrinos with masses below or at the eV scale. 
Such dark radiation is parameterized as the number of effective thermalized neutrino species, $N_{eff}$, and impacts several cosmological events 
including nucleosynthesis and the time of matter-radiation equality. In this section we show the constraints on the parameters of the model in order to satisfy the cosmological bounds.

The contribution of the very light sterile neutrinos to $N_{eff}$ depends on how they have been thermalized. In this case the thermalization can take place through two mechanisms: \\

\begin{itemize}

\item Through the sterile-active oscillations, see for example Refs.~\cite{Barbieri:1989ti,Enqvist:1990dq,Hannestad:2012ky,Mirizzi:2012we} for different studies.

\item Through new interactions, see for example the studies~\cite{SolagurenBeascoa:2012cz,Anchordoqui:2012qu,Perez:2013kla}.

\end{itemize}

In the models proposed above we predict that the sterile neutrinos must have mass below or at the scale similar to the left-handed neutrinos 
and the mixing angles between the left and right-handed neutrinos are not predicted. Assuming that the mixing angle is very small we 
investigate the bounds from the measured $N_{eff}$ values on the new interactions as in the second mechanism mentioned above. 
The change on $N_{eff}$ due to the contribution of the light right-handed neutrinos is given by
\begin{equation}
\Delta N_{eff}= N_{eff} - N_{eff}^{SM}=3 \left( \frac{T^{N}_{dec}}{T^{\nu_L}_{dec}} \right)^4 = 3 \left( \frac{g (T^{\nu_L}_{dec})}{g(T^{N}_{dec})} \right)^{4/3},
\end{equation}
where $g(T)$ is the effective number of degrees of freedom at temperature $T$, $N_{eff}^{SM}=3.045$ is the contribution of the SM neutrinos and $T^{\nu_L}_{dec}=3$ MeV 
is their decoupling temperature. In this article we will use the following bounds on $N_{eff}$ reported in the recent analysis in Ref.~\cite{Bernal:2016gxb}:
\begin{eqnarray}
\Delta N_{eff} &<& 0.28  \  {\textrm{when}} \  H_0=68.7^{+0.6}_{-0.7}  \  \textrm{Mpc}^{-1} \textrm{km/s}, \\
\Delta N_{eff} &<& 0.77 \  {\textrm{when}}  \ H_0=71.3^{+ 1.9}_{-2.2} \  \textrm{Mpc}^{-1} \textrm{km/s}.
\end{eqnarray} 
These bounds have been obtained using different data set, for details about these bounds see Ref.~\cite{Bernal:2016gxb}.

In order to constrain the new interactions present in our model we use these bounds and evaluate 
the decoupling temperature for different values of the input parameters $g_{BL}$ and $M_{Z_{BL}}$.

The decoupling temperature of the right-handed neutrinos can be computed using the relation
\begin{equation}
\Gamma_{N}(T^{N}_{\text{dec}}) = H(T^{N}_{\text{dec}}),
\label{dec}
\end{equation}
where the annihilation rate of right-handed neutrinos with other SM particles is given by
%
\begin{equation}
\begin{split}
& \Gamma_{N} (T)=n_{N} (T) \sum_f \avg{\sigma_f ( N N \to \bar{f} f )v} \\
&= \sum_f \frac{g_{N}^2}{n_{N}} \int{\frac{d^3p}{(2\pi)^3} f_N (p)} \int{\frac{d^3 k}{(2\pi)^3} f_N (k)}  \sigma_f(s) v_M.\\
\end{split}
\end{equation}
%
Here, $v_M$ represents de Moller velocity $v_M=(1-\cos \theta)$ where $\theta$ is the angle between the two colliding particles. 
The function $f_N (k)$ is the Fermi-Dirac distribution, defined as
\begin{equation}
f_N (k) =\frac{1}{e^{k/T}+1},
\end{equation}
and the number density of the right-handed neutrinos, $n_N$, which spin number is $g_{N}=2$, is given by
\begin{equation}
n_{N}=g_{N}\int{\frac{d^3 k}{(2\pi)^3} f_N (k)}= \frac{3\xi (3) T^3}{2\pi^2}.
\end{equation}
The cross-section of the right-handed neutrinos annihilation into SM particles is given by
\begin{equation}
\sigma_f(s)= \frac{g_{BL}^4}{12 \pi} \frac{ N_c^f (Q_{BL}^f)^2 s}{ \left[ (s- M_{Z_{BL}}^2)^2 + M_{Z_{BL}}^2 \Gamma_{Z_{BL}}^2 \right]},
\end{equation}
where $s=2pk (1-\cos \theta)$, $Q_{BL}^f$ is the $B-L$ charge of the SM fermions, $-1$ for leptons and $1/3$ for quarks, and 
$N_c^f$ is 3 for quarks and 1 for leptons. Now, working in the relevant limit $M_{Z_{BL}}^2 \gg s$ one finds
\begin{eqnarray}
\Gamma_N (T)&=& \frac{49 \pi^5 T^5}{194400 \ \xi(3)} \left( \frac{g_{BL}}{M_{Z_{BL}}}\right)^4 \sum_f Q_{BL}^f N_c^f. 
\end{eqnarray}
On the other hand, we have the Hubble parameter, defined as
\begin{equation}
H(T)= \sqrt{\frac{8\pi G_N \rho(T)}{3}}=\sqrt{\frac{4\pi^3G_N ( g(T) + \frac{21}{4}) }{45}}T^2,
\end{equation}
where g(T) represents the relativistic degrees of freedom of the SM which values are given in Ref.~\cite{SolagurenBeascoa:2012cz}.
Therefore, now we are ready to understand the cosmological constraints in this model.
\begin{figure}[t]
\includegraphics[width=0.99\linewidth]{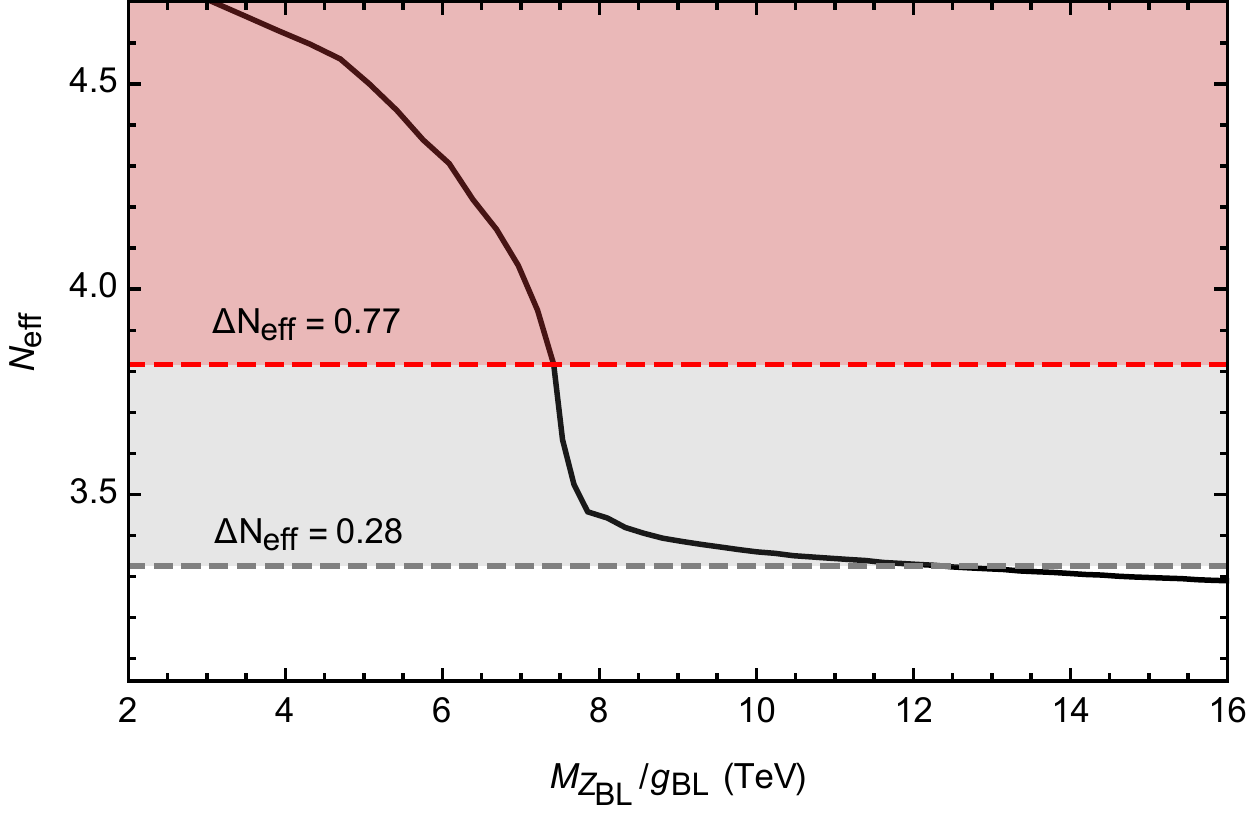}
\caption{Effective number of relativistic degrees of freedom vs. the ratio of the $B-L$ gauge boson mass and gauge coupling. 
The horizontal lines correspond to the upper bounds mentioned in the text and reported in Ref.~\cite{Bernal:2016gxb}.}
\label{NeffMZBLggs}
\end{figure}

In Fig. 3 we show the numerical results for the effective number of relativistic degrees of freedom as a function of the ratio between 
the $ B-L$ gauge boson mass and gauge coupling. As one can appreciate, the ratio $M_{Z_{BL}}/g_{BL}$ must be larger than $7-8$ TeV 
in order to be in agreement with the cosmological constraints. This bound is competitive with the collider and electroweak precision bounds 
$M_{Z_{BL}}/g_{BL} > 6-7$ TeV~\cite{Carena:2004xs,Cacciapaglia:2006pk,Salvioni:2009jp}. In this way we 
show that one can have a consistent picture with cosmology in these models even if the right-handed neutrinos are very light.

\FloatBarrier
{\section{V. Lepton Number Violating Processes}}
%
\begin{figure}[h]
\includegraphics[width=0.49\linewidth,keepaspectratio=true,clip=true]{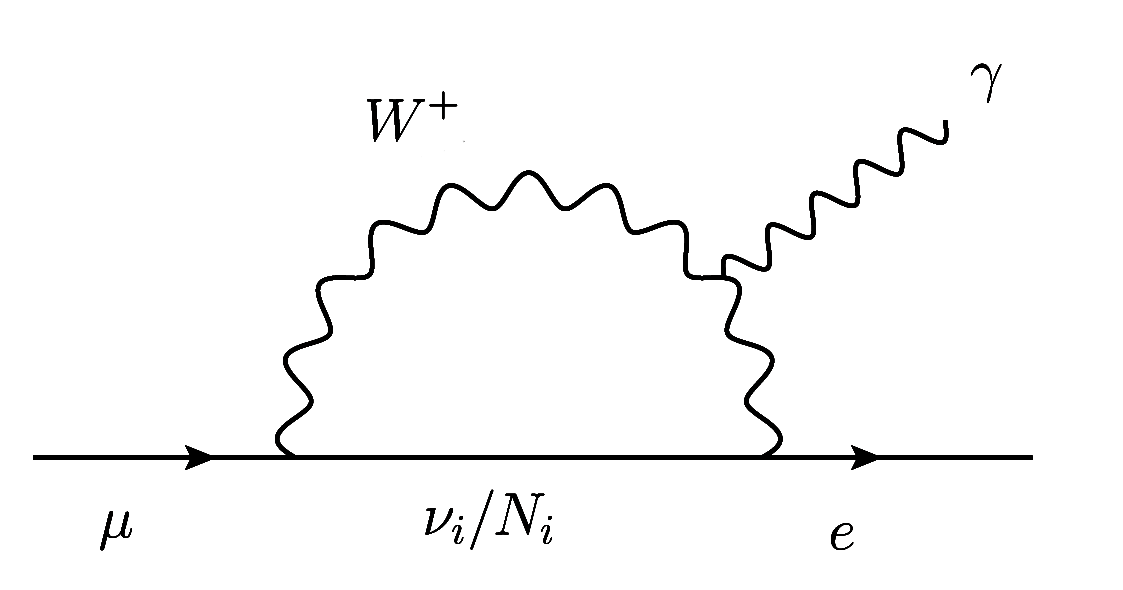}
\includegraphics[width=0.49\linewidth,keepaspectratio=true,clip=true]{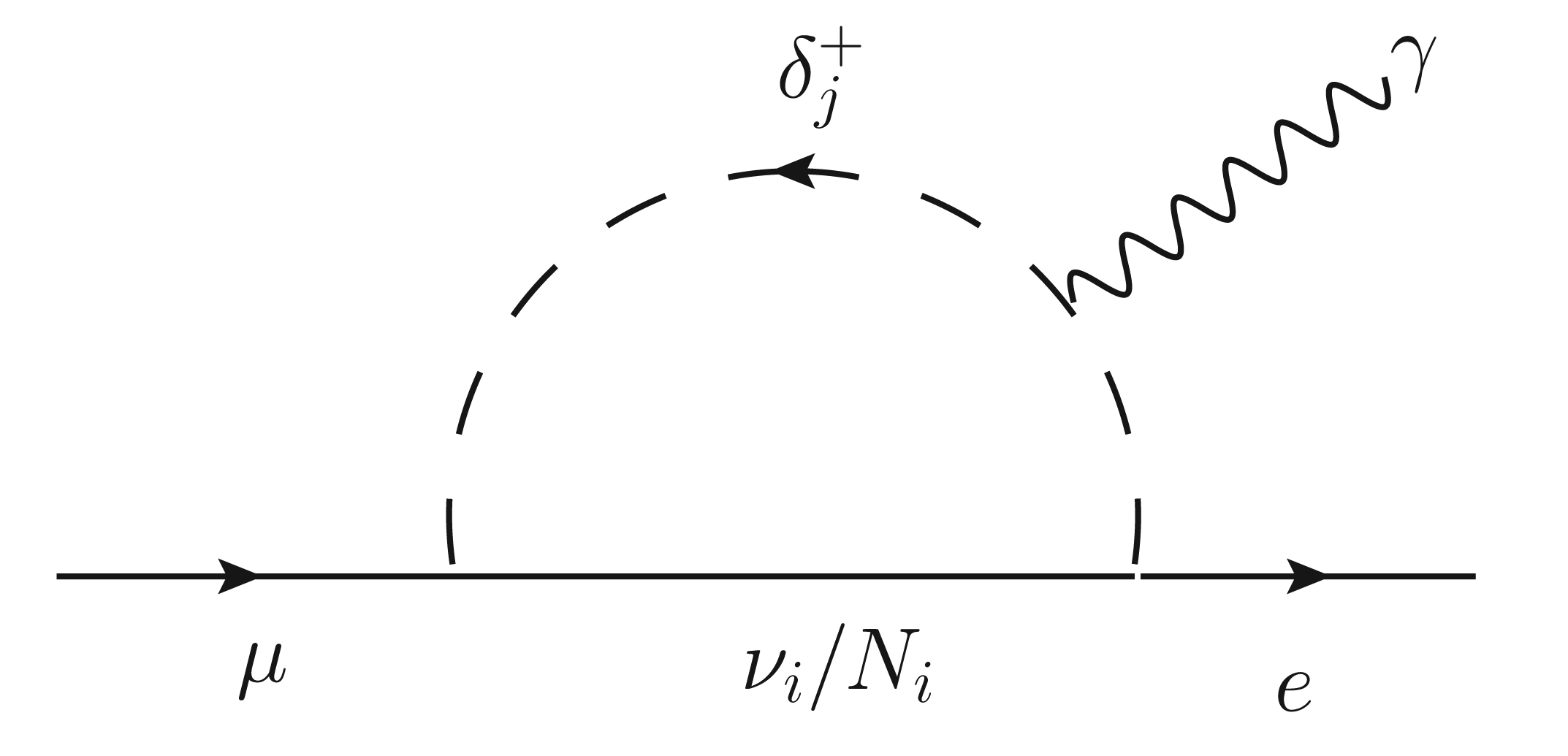}
\caption{Processes contributing to $\mu \to e \gamma$ in the Radiative Seesaw.}
\label{muegamma}
\end{figure}
In the canonical $B-L$ models for neutrino masses the lepton number violating processes such as $\mu \to e \gamma$ are highly suppressed.
The branching ratio, see left graph in Fig.~\ref{muegamma} for the Feynman graph, is strongly suppressed by unitarity constraints 
on the mixing matrices $V_\nu$, as well as in the SM case, which is given by
 \begin{equation}
A_R^{W}\approx \frac{g_L^2 \ e}{64 \pi^2}\frac{m_\mu}{m_{W}^2}\sum_i(V_\nu)_{e i}(V_\nu^*)_{\mu i}F\left(\frac{m_{\nu_i}^2}{m_{W}^2}\right).
\end{equation}
where $V_\nu$ refers to the rotation matrix which brings neutrinos to their flavor-diagonal basis. 
Here one can see how the loop factor, defined as 
\begin{equation}
F(x)=\frac{10-43\,x+78\,x^2-49\,x^3+18\,x^3\, \text{Log}(x)+4\,x^4}{6(1-x)^4}
\end{equation}
gives a constant for a very suppressed mass ratio, which is therefore suppressed by the unitarity constraints on the mixing matrix.
However, in the context of the Radiative Seesaw mechanism, the presence of the Yukawa couplings $\lambda_L$ and $\lambda_R$, which enter in the amplitude according to right graph in Fig.~\ref{muegamma}, avoid the unitarity suppression and the amplitudes of the process $\mu \to e \gamma$ mediated by $\delta^+$ are given by 
\begin{eqnarray}
 A_L^{\delta^+} &=& \frac{e}{8\pi^2}\frac{m_\mu}{m_{\delta^+}^2}\sum_{c,d}(\lambda_R^*)^{ce}\lambda_R^{d\mu}\sum_iV_N^{ci}(V_N^*)^{di}\,G\left(
 \frac{m_{N_i}^2}{m_{\delta^+}^2}\right), \nonumber \\ 
 \\
 A_R^{\delta^+} &=& \frac{e}{4\pi^2}\frac{m_\mu}{m_{\delta^+}^2}\sum_{c,d}(\lambda_L^*)^{ce}\lambda_L^{d\mu}\sum_i(V_\nu^*)^{ci}V_\nu^{di}\,G\left(
 \frac{m_{\nu_i}^2}{m_{\delta^+}^2}\right),  \nonumber \\
\end{eqnarray}
where the loop-factor is given by 
\begin{equation}
G(x)=\frac{1-6x+3x^2+2x^3-6x^2\text{Log}(x)}{12(1-x)^4}.
\end{equation}
Here we have neglecting the mixing among charged scalars. In the above expressions $A_R$ and $A_L$ refer to the amplitudes entering in the process $\mu \to e \gamma$ according to
\begin{equation}
  \mathcal{A}(\mu\to e \gamma)=i\overline{u_e}(p-q)\epsilon^*_{\nu}\sigma^{\nu \mu}q_{\mu}[A_RP_R+A_LP_L]u_{\mu}(p),
\end{equation}
where $p^{\mu}$ and $q^\mu$ are the muon and photon quadrimomenta.
In Fig.~5 we show the predictions for the branching ratio for the process $\mu \to e \gamma$. 
\begin{figure}
 \includegraphics[width=0.9\linewidth,keepaspectratio=true,clip=true]{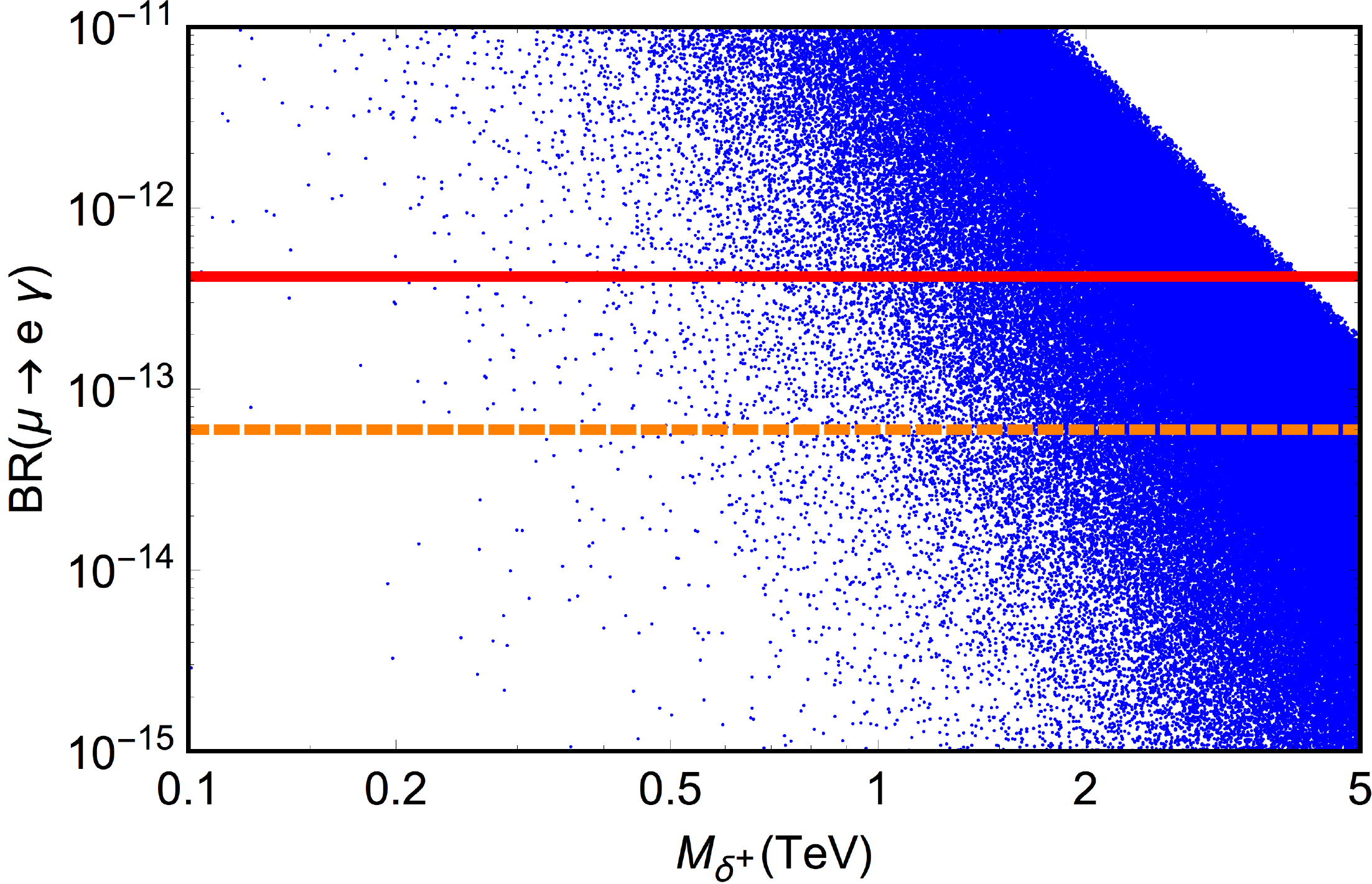}
 \label{Br}
 \caption{Prediction on the branching ratio of the process $\mu \to e \gamma$ as a function of $M_{\delta^+}$. Here, the Yukawa couplings $\lambda_L$ and $\lambda_R$ range from $10^{-4}$ to $10^{-1}$. The red line shows the current experimental upper bound on $\mu \to e \gamma$, $4.2 \times 10^{-13}$~\cite{Adam:2013mnn} and the orange dashed line shows the projected bound $6\times 10^{-14}$~\cite{Baldini:2013ke}.}
\end{figure}
As Fig.~5 shows, for a reasonable light choice of the mass of the charged singlet, we have LFV signals entering in our range of visibility, which makes the model testable regarding current and future experiments. In Fig.~5, the red line shows the current upper bound on $\mu \to e\gamma$, given by the MEG experiment at PSI,
\begin{equation*}
\rm{Br}(\mu \to e \gamma)< 4.2 \times 10^{-13} ~\mbox{\cite{Adam:2013mnn}},
\end{equation*}
which is expected to be further improved to $6\times 10^{-14}$ (orange dashed line) \cite{Baldini:2013ke}.
Even if $\mu \to e \gamma$ is nowadays the most constrained LFV process, it is very interesting to 
also look at $\mu \to e$ conversion, which bounds on different nuclei are reported in Table~\ref{bounds}.
Projected bounds on $\mu-e$ conversion, unlike on $\mu \to e \gamma$, are expected to be improved up to four orders of magnitude according to future experiments such as DeeMe at J-PARC \cite{Aoki:2010zz}, with a sensitivity of $10^{-14}$, COMET at J-PARC \cite{Cui:2009zz}, with $10^{-16}$, and Mu2e at Fermilab \cite{Morescalchi:2016uks}, with $6\times 10^{-17}$. 
Therefore, these projected bounds provide a good motivation for the study of these processes.

  \begin{table}[h]
\caption{Current bounds on LFV for $\mu \to e \gamma$ and $\mu-e$ conversion in different nuclei.}
\begin{tabular}{ c | r  }
LFV process & Current bounds $ \,\,\,\,\,\,$\\
\hline
$\,\,\,\,\,\,\rm{Br}(\mu \to e \gamma)\,\,\,\,\,\,$ & $4.2 \times 10^{-13} ~\mbox{\cite{Adam:2013mnn}}$ \\
 $\displaystyle \rm{Br}(\mu~\text{Ti} \to e~\text{Ti})$ & $\,\,\,\,\,\,4.3 \times 10^{-12} ~ \mbox{\cite{Dohmen:1993mp}}$ \\
$\displaystyle \rm{Br}(\mu~\text{Au} \to e~\text{Au})$  &  $7 \times 10^{-13}  ~ \mbox{\cite{Bertl:2006up}}$ \\
 $\displaystyle \rm{Br}(\mu~\text{Pb} \to e~\text{Pb})$ & $4.6\times 10^{-11} ~ \mbox{\cite{Honecker:1996zf}}$ \\
 \hline
 \end{tabular}
 \label{bounds}
 \end{table}
 
\begin{figure}[h]
 \includegraphics[width=0.9\linewidth,keepaspectratio=true,clip=true]{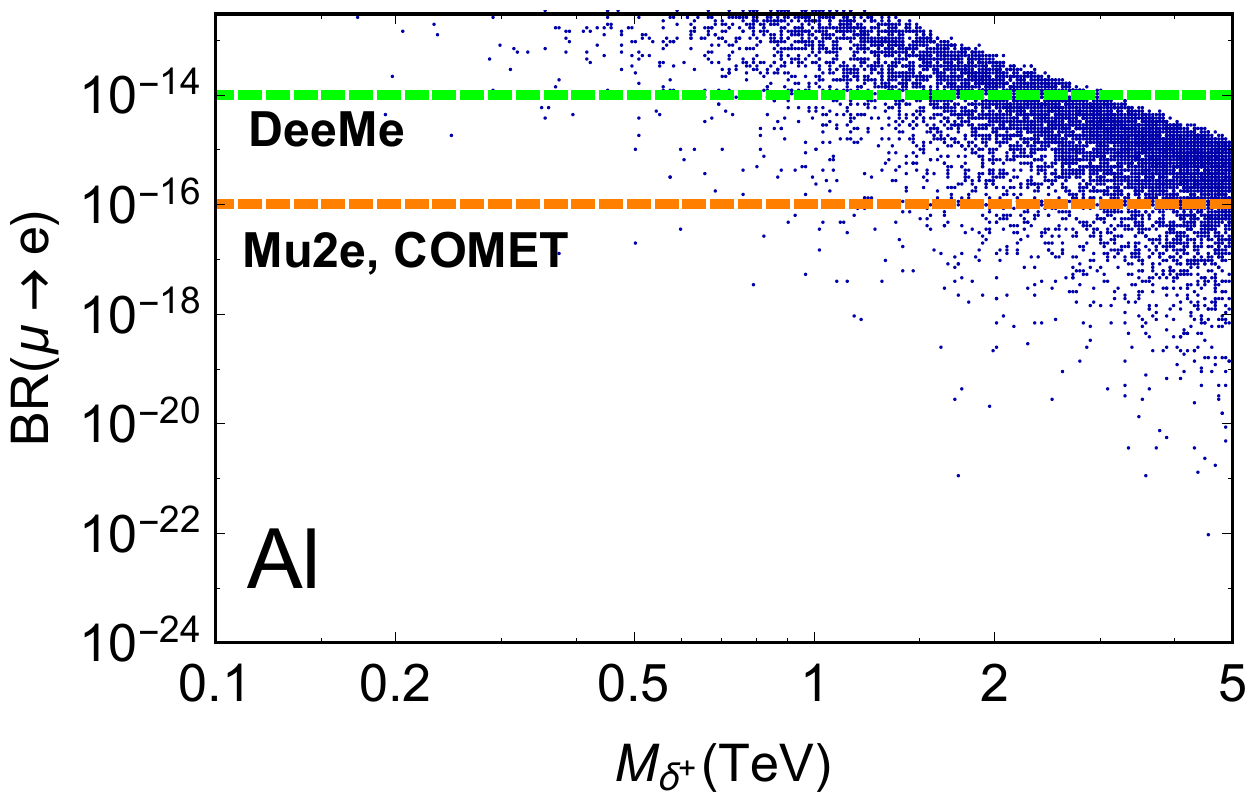}
\includegraphics[width=0.9\linewidth,keepaspectratio=true,clip=true]{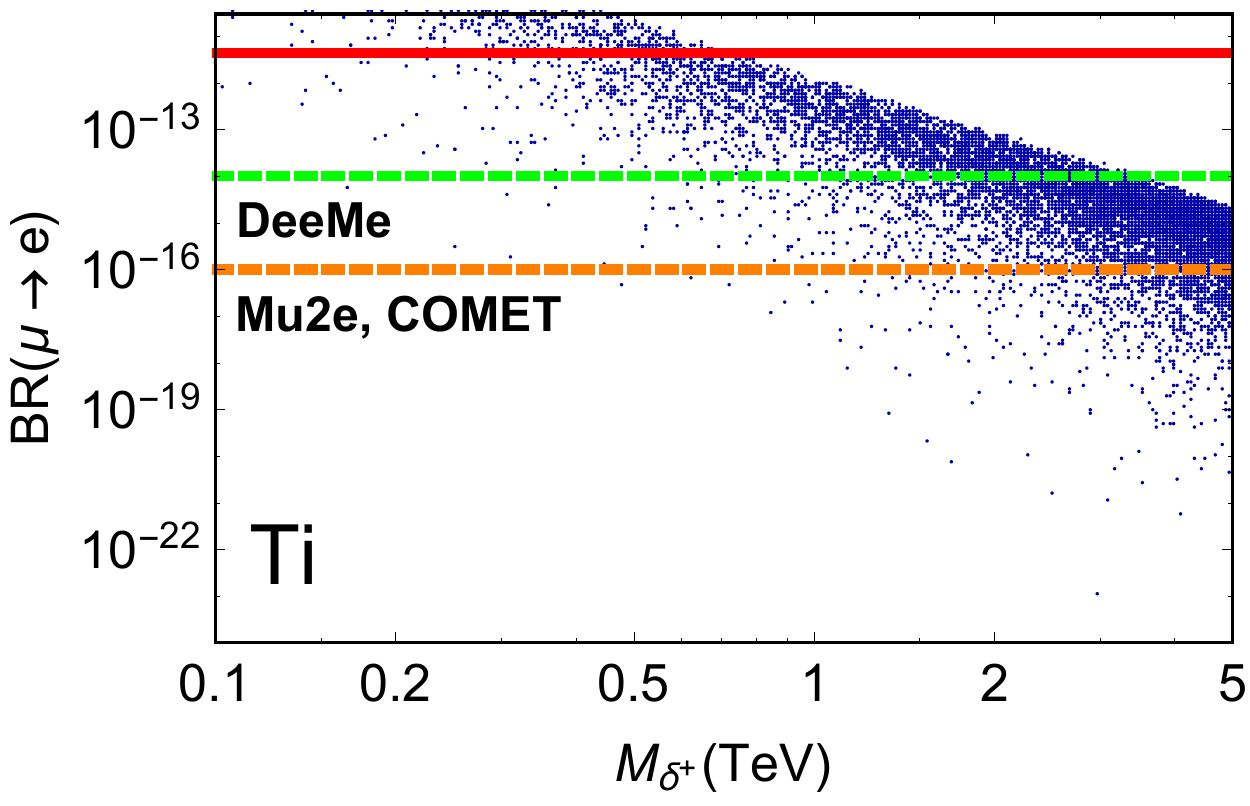}
 \label{mue}
 \caption{Prediction on the branching ratios of the process $\mu-e$ conversion in Al and Ti nuclei as a function of the mass of the singly charged Higgs, $M_{\delta^+}$. Here, the Yukawa couplings $\lambda_L$ and $\lambda_R$ entering in the process range from $10^{-4}$ to $10^{-1}$. The solid red lines show the experimental upper bounds on $\mu-e$ for the different nuclei (see Table~\ref{bounds}). The dashed lines show the projected sensitivities of the experiments DeeMe with $10^{-14}$~\cite{Aoki:2010zz} (green line), COMET with $10^{-16}$ and Mu2e with $6\times 10^{-17}$~\cite{Cui:2009zz,Morescalchi:2016uks} (orange line).}
\end{figure}

\begin{figure}[h]
 \includegraphics[width=0.9\linewidth,keepaspectratio=true,clip=true]{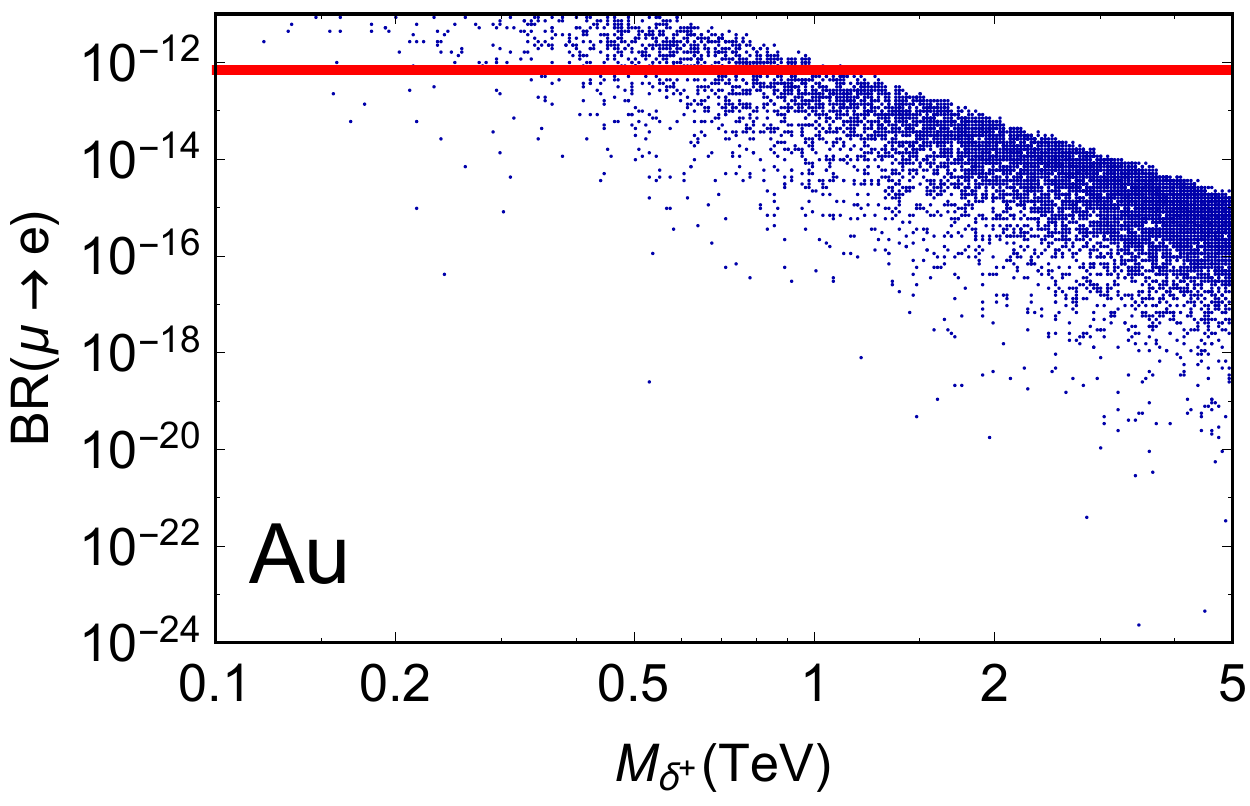}
 \includegraphics[width=0.9\linewidth,keepaspectratio=true,clip=true]{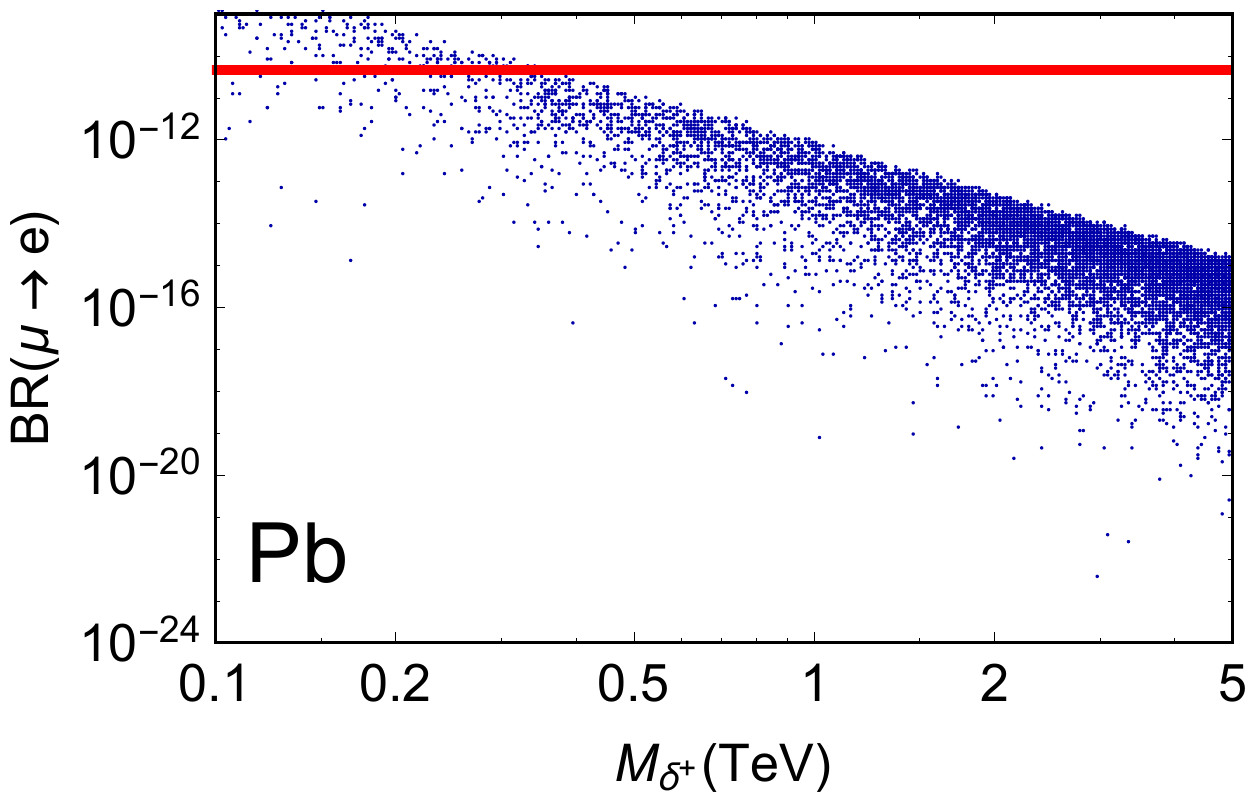}
 \label{mue}
 \caption{Prediction on the branching ratios of the process $\mu-e$ conversion in Au and Pb nuclei as a function of the mass of the singly charged Higgs, $M_{\delta^+}$. Here, the Yukawa couplings $\lambda_L$ and $\lambda_R$ entering in the process range from $10^{-4}$ to $10^{-1}$. The solid red lines show the experimental upper bounds on $\mu-e$ for the different nuclei (see Table~\ref{bounds}). 
 }
\end{figure}

In Figs.~6 and 7 we show the predictions on the branching ratio for $\mu-e$ conversion processes in nuclei such as aluminium (Al), titanium (Ti), gold (Au) and lead (Pb). The computation of the branching ratios for $\mu-e$ conversion has been done following Ref.~\cite{FileviezPerez:2017zwm} (see this reference for details). For the first two nuclei, the projected bounds are $10^{-14}$ (DeeMe)~\cite{Aoki:2010zz} and $10^{-17}$ (COMET and Mu2e)~\cite{Cui:2009zz,Morescalchi:2016uks}, as we show in the dashed lines. \\

We would like to emphasize on the testability of the Radiative Seesaw mechanism model via LFV signals like $\ell_i \to \ell_j \gamma$ and $\mu-e$ conversion in comparison with other models, like the canonical Type I seesaw mechanism, which are hopeless to be tested in the current and even future experiments. Apart from the prediction of light Sterile neutrinos, the fact of predicting accessible LFV at colliders makes the model one of the most attractive $B-L$ extensions of the Standard Model. 

{\section{VI. Summary}}
%
We have discussed the relation between the generation of neutrino masses and the spontaneous breaking of the $B-L$ gauge symmetry.
We have proposed two simple models where the neutrino masses are generated dynamically in the context of theories where the $B-L$ gauge symmetry is spontaneously broken.
In the first model the $B-L$ symmetry is broken in two units but the right-handed neutrinos are predicted to be very light; they must have masses below 
the eV scale. In this case the neutrino masses are generated through the $B-L$ Inverse Type II seesaw mechanism.
In the second model the Majorana masses for the right-handed and the SM neutrinos are generated at the quantum level through the $B-L$ radiative mechanism. 
The right handed neutrino masses are predicted to be very light as in the first model. Only the $B-L$ radiative seesaw mechanism can be realistic without assuming small gauge coupling 
and could be tested in the near future.  	

We have discussed the main phenomenological and cosmological constraints. The bounds coming from the constraints on the effective number 
of relativistic degrees of freedom have been discussed in detail. These bounds are as competitive as the collider bounds on the $B-L$ breaking scale. 
The implications for the decays of the $B-L$ gauge boson have been discussed in order to understand the testability of these models at collider experiments. 
We have investigated in detail the predictions for lepton number violating processes such as $\mu \to e \gamma$ and $\mu \to e$ conversion in nuclei, 
showing that the radiative $B-L$ seesaw mechanism could be tested in the future LFV experiments.
The $B-L$ radiative seesaw mechanism proposed in this Letter can be considered as an appealing mechanism for neutrino masses. 

\textit{Acknowledgments}: P. F. P. thanks Mark B. Wise for discussions and comments on the manuscript. 
The work of P.F.P. has been supported by the U.S. Department of Energy under contract No. DE-SC0018005.
C.M. thanks Toni Pich for discussions. The work of C.M. has been supported in part by the Spanish Government and ERDF funds from the EU Commission [Grants No. FPA2014-53631-C2-1-P and SEV-2014-0398] and ``La Caixa-Severo Ochoa" scholarship.


\end{document}